\def\refeq#1{(\ref{#1})}
\def\sh{\sinh\frac\pi{\pi-\gamma}}
\def\cth{\coth\frac\pi{\pi-\gamma}}
\documentstyle[sprocl,amsmath,amsfonts,amssymb,graphicx]{article}
\begin{document}

\newcommand{\e}{{\rm e}}
\renewcommand{\L}{\ln\Lambda}
\newcommand{\lam}{\lambda}
\newcommand{\Tr}{{\rm Tr}}

\title{THE SPIN-1/2 $XXZ$ CHAIN  AT FINITE MAGNETIC FIELD: CROSSOVER 
  PHENOMENA DRIVEN BY TEMPERATURE}

\author{A. Kl\" umper}
\address{Universit\"at Dortmund, Fachbereich Physik, Otto-Hahn-Str. 4\\
D-44221 Dortmund, Germany}
\author{J.R. Reyes Mart\'\i nez, C. Scheeren}
\address{Universit\"at zu K\"oln,
        Institut f\"ur Theoretische Physik,
        Z\"ulpicher Str. 77\\
        D-50937 K\"oln, Germany}
\author{M. Shiroishi}
\address{Institute for Solid State Physics, University of Tokyo,\\
 Kashiwanoha, Kashiwa-shi, Chiba 277-8581,Japan}

\maketitle

\vskip0.5cm
\centerline{Dedicated to Professor Rodney Baxter on the occasion
of his 60th birthday}
\vskip0.5cm

\begin{abstract}
We investigate the asymptotic behaviour of spin-spin correlation
functions for the integrable Heisenberg chain. To this end we use the
Quantum Transfer Matrix (QTM) technique developed in \cite{AK} which
results in a set of non-linear integral equations (NLIE). In the case
of the largest eigenvalue the solution to these equations yields the
free energy and by modifications of the paths of integration the
next-leading eigenvalues and hence the correlation lengths are
obtained.  At finite field $h>0$ and sufficiently high temperature $T$
the next-leading eigenvalue is unique and given by a 1-string solution
to the QTM taking real and negative values thus resulting into exponentially
decaying correlations with antiferromagnetic oscillations. At
sufficiently low temperatures a different behaviour sets in where the
next-leading eigenvalues of the QTM are given by a complex conjugate pair
of eigenvalues resulting into incommensurate oscillations.

The above scenario is the result of analytical and numerical
investigations of the QTM establishing a well defined crossover
temperature $T_c(h)$ at which the 1-string eigenvalue to the QTM gets
degenerate with the 2-string solution.  Among other things we find a
simple particle-hole picture for the excitations of the QTM allowing
for a description by the dressed charge formulation of CFT. 
\end{abstract}

\vfill

\noindent
{\bf Keywords:} exact solution, quantum transfer matrix, strongly correlated
systems, crossover phenomena, level crossing, conformal field theory


\section{Introduction} 
\label{sec:1}

In this contribution we report on a new crossover phenomenon observed
in the longitudinal correlation function at finite magnetic field
$h$ and finite temperature $T$ of the $XXZ$ spin-chain with 
anisotropy parameter $0<\Delta\le 1$. 

As is well known any thermodynamic quantity derived from the free
energy of the one-dimensional $XXZ$ model is an analytic function of
finite $h$ and $T$. Phase transitions and associated mathematical
singularities may and do occur in the ground state, i.e. at
$T=0$. However, quantities not obtained from the free energy, but
characterising the asymptotics of correlation functions may show
``their own'' non-analyticities {\em at finite} temperatures. Indeed
the $XXZ$ chain in an external magnetic field does show well defined
non-analyticities in the {\em correlation lengths} of the longitudinal
spin-spin correlation functions. We want to point out that these
crossover phenomena are a result of strong correlations and finite
temperature. The crossover does not take place for the case $\Delta=0$
corresponding to free fermions.

For the attractive regime $-1<\Delta\le 0$ of the $XXZ$ chain the
investigations are not yet carried out for finite field $h$ and finite $T$.
Here the physical phenomena are expected to be much richer. Even for
the case of vanishing magnetic field $h=0$ a sequence of crossovers
(commensurate-incommensurate-commensurate) for an increase of $T$ from 0
to $\infty$ was found \cite{FKM}.

The method we apply (as that used in \cite{FKM}) is based on a mapping
of the $XXZ$ chain at finite temperatures onto a classical model in
two dimensions and the formulation of a suitable quantum transfer
matrix (QTM) describing the transfer along the chain. The spectrum of
the QTM succumbs to a Bethe ansatz (BA) treatment and the
corresponding BA equations can be cast into the form of non-linear
integral equations. Crossover phenomena of correlation lengths
manifest themselves as level crossings of next-leading eigenvalues of
the QTM.  At finite field $h>0$ and sufficiently high temperature the
relevant eigenvalues for the longitudinal correlation functions are
1-string and 2-string solutions (both solutions belong to the $S = 0$
sector of the model). The truely next-leading eigenvalue is unique
and given by the 1-string solution to the QTM taking real and negative
values thus resulting into exponentially decaying correlations with
antiferromagnetic oscillations. 
In some sense at sufficiently high temperature the properties of the
system are determined by the longitudinal (``classical'') terms of the 
Hamiltonian, i.e. the $S^zS^z$ coupling and the field in $z$ direction,
which dominate over the transversal exchange (``quantum mechanical'') terms.
At sufficiently low temperature a
different behavior is expected on grounds of predictions by conformal
field theory (CFT)\cite{car,affb}. In particular, correlations with
incommensurate $2k_F$ oscillations are expected\cite{BogK89,KBI}. As a
consequence of this, the QTM has to develop complex conjugate pairs of
eigenvalues at sufficiently low temperatures. This scenario has not
been investigated before. The purpose of this contribution is to
present the first ``Bethe ansatz'' study of this crossover phenomenon
and to provide reasonably accurate values for the crossover temperature
$T_c$.
As pointed out already, for the free fermion case $T_c=\infty$, i.e.
the crossover does not take place. In physical terms this may be understood
in the way that due to the absence of longitudinal couplings the 
longitudinal terms never dominate over the transversal terms.

This report is organized as follows. In Sec.~2 we introduce some basic
definitions of the $XXZ$ chain and present its properties at low
temperature as obtained within conformal field theory (CFT).  In
Sec.~3 the approach to thermodynamic properties by use of a lattice
path integral formulation and the quantum transfer matrix (QTM) is
reviewed.  In particular the set of non-linear integral equations for
the two auxiliary functions ${\mathfrak{a}}(x)$ and
$\overline{\mathfrak{a}}(x)$ corresponding to the energy density
functions of spinons with spin $\pm 1/2$ are given.  In Sec.~4
numerical results are given for the correlation length and Fermi
momentum of the longitudinal spin-spin correlation function. Finally,
the particle-hole picture resulting at low temperatures is discussed.
A complete exposition of how this is related to the dressed charge
formulation of CFT will appear in \cite{KMSS}.

\section{Anisotropic Heisenberg model}
\label{sec:xxz}

The $XXZ$ model is defined by the Hamiltonian
\begin{equation}
H= J\sum_{<i,j>}\left[S^x_iS^x_j+S^y_iS^y_j+\Delta S^z_iS^z_j\right]
-h\sum_{j} S^z_j,
\end{equation}
where $J > 0$, $h > 0$ and $S^{x,y,z}$ are spin-1/2 operators.  For
$T=0$ the system is in one of three phases. For anisotropy parameter
$|\Delta| > 1$ we have two ordered phases and for $|\Delta| \leq 1$ a
critical phase. In the latter case
the anisotropy parameter is conveniently parameterized by
$\Delta = \cos\gamma$. In the following we take 
$0< \gamma < \pi/2$ (repulsive regime) for simplicity. The excitations
are gapless spinons with spin $1/2$
\begin{equation}
\epsilon_F(k)=v \sin k,\qquad 0\le k\le\pi,
\label{spinons}
\end{equation}
\begin{equation}
v=\frac{\sin\gamma}{\gamma} \pi J.
\end{equation}
At zero temperature the spin correlations decay
algebraically\cite{yya,yyb,yyc,lp,fog}, with exponents depending on
the magnetic field in a nontrivial way
\cite{KBI}. At finite temperature the correlations decay exponentially 
$\e^{-r/\xi}$ with in general different correlation lengths $\xi$ for
different correlation functions. At low temperatures the correlation
lengths can be related by conformal mappings to the scaling dimensions
$x$ of the fields at $T=0$
$$
\xi = \frac{v}{2\pi x T}\hbox{ for}\quad T\ll 1
$$
The scaling dimensions $x$ are calculated from finite size
corrections to the energy levels of the Hamiltonian and scaling predictions
by CFT
\begin{equation}
E_x-E_0=\frac{2\pi}{L}v(x+N^++N^-) + o\left( \frac 1 L\right),
\end{equation}
where $N^+$ and $N^-$ are integers labelling the conformal tower.
The critical exponents of the spin-1/2 $XXZ$ model are those of a $c = 1$
Gaussian theory.  At zero magnetic field the exponents are given in terms
of the anisotropy parameter $\gamma$
\begin{equation}
x=\frac{1-\gamma/\pi}{2}S^2+\frac{1}{2(1-\gamma/\pi)}m^2,
\end{equation}
where $S$ and $m$ are integers corresponding to the $S^z$ component of the 
state and the number of excitations from the left to the right Fermi point,
respectively.
The longitudinal correlation is
\begin{equation}
\label{corr}
\langle{ S^z_0}{ S^z_r}\rangle\simeq C_1\frac{\cos(2k_Fr)}
{r^{1/(1-\gamma/\pi)}}-\frac{C_2}{r^2},
\end{equation}
where $C_1$ and $C_2$ do not depend on the distance $r$. For zero magnetic
field the Fermi momentum $k_F$ is equal to $\pi/2$.
For finite magnetic field and temperature we expect a deviation of $x$
and $k_F$ from the above quoted values.


\section{Finite Temperatures}
\label{sec:temp}

The properties of the quantum system at finite temperature are
determined within a path integral formulation. The partition function
of the 1d quantum system at finite temperature is mapped to that of
a 2d classical system pictorially represented by\\

\unitlength1cm
\begin{picture}(0,3.5)
\vbox{\hbox{$\Tr\e^{-\beta H}=$part.fct.}
\hbox{}\hbox{}\hbox{}}
\includegraphics[width=6cm]{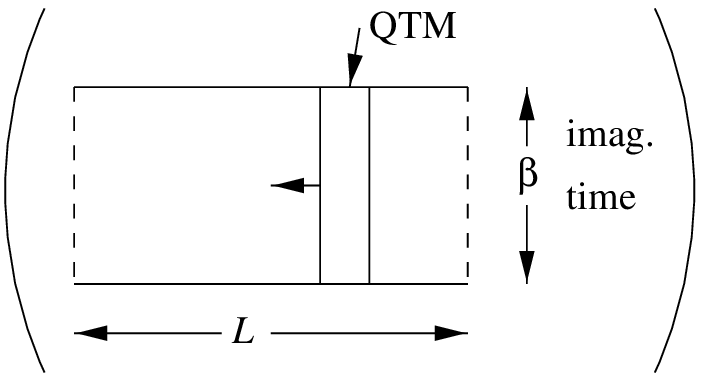}
\vspace{1cm}
\end{picture}

\begin{figure}
\centering
\includegraphics[width=0.8\textwidth]{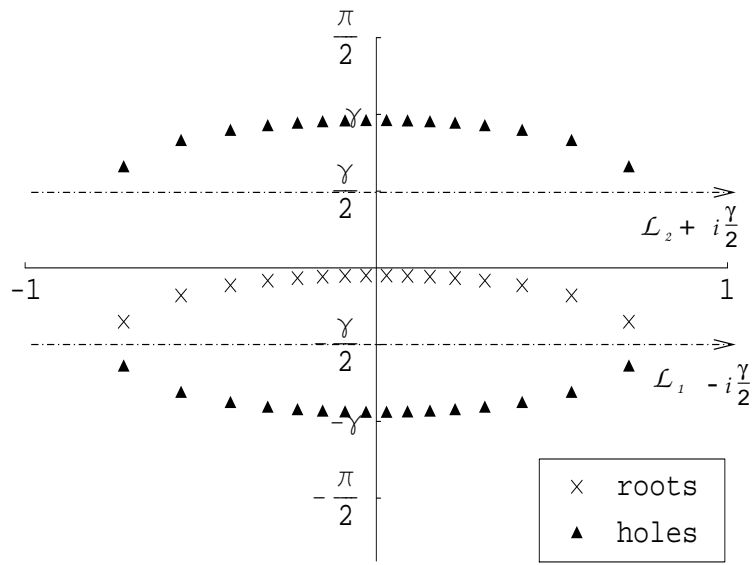}
\caption{Distribution of roots and holes for the largest eigenvalue
for anisotropy parameter $\gamma=\pi/3$, reciprocal temperature
$\beta= 12$ and $h = 0.6$. Note that straight contours with imaginary
parts $\pm\gamma/2$ separate the roots from the holes. The contours
${\mathcal{L}}_{1}- i\gamma/2$ and 
${\mathcal{L}}_{2}+ i\gamma/2$ are depicted by 
dashed-dotted lines. Note that roots and holes get arbitrarily close to
${\mathcal{L}}_{1}- i\gamma/2$
for low temperatures $T$ and positive $h$.}
\label{fig:gs}
\end{figure}

\noindent
For the classical system we can choose a suitable transfer
approach. The quantum transfer matrix (QTM), i.e. the column-to-column
transfer matrix, possesses a unique largest eigenvalue and a spectral
gap to the next-leading eigenvalues which persists even in the limit
of infinite Trotter number as long as the temperature is
finite\cite{Suz85,SuIn87,Suz90,AK,Kun,Sak} . Hence the free-energy and
correlation lengths are simply determined from the largest $\Lambda_0$
and next largest eigenvalues $\Lambda_j$ of the QTM
\begin{eqnarray}
f&=&-\frac{1}{\beta}\lim\L_0\nonumber\\
C(r)&=&A_1 \left(\frac{\Lambda_1}{\Lambda_0}\right)^r
+A_2 \left(\frac{\Lambda_2}{\Lambda_0}\right)^r
+... .
\label{expansion}
\end{eqnarray}
with coefficients $A_1$, $A_2$ given by matrix elements.
The diagonalization of the QTM can be performed by means of a 
Bethe ansatz resulting into eigenvalue equations of the
form of Baxter's $\Lambda-q$ equation. For finite Trotter number $N$
we find
\begin{equation}
\Lambda(x)q(x)=\e^{-\beta h/2}\Phi(x-i\gamma/2)q(x+i\gamma)
+\e^{+\beta h/2}\Phi(x+i\gamma/2)q(x-i\gamma)
\end{equation}
where 
\begin{eqnarray}
\Phi(x)&=&[\sinh(x-ix_0)\sinh(x+ix_0)]^{N/2},\qquad x_0:=\frac\gamma 2-
\frac\beta N\cr
q(x)&=&\prod_j\sinh(x-x_j)
\end{eqnarray}
The roots $x_j$ are determined from the Bethe ansatz equation
\begin{equation}
p(x_j)=-1, \quad 
\hbox{where}\quad  p(x):=\e^{-\beta h}\frac{\Phi(x-i\gamma/2)q(x+i\gamma)}
{\Phi(x+i\gamma/2)q(x-i\gamma)}
\end{equation}
thus rendering $\Lambda(x)$ analytic. We like to note that in general
there are more solutions to the Bethe ansatz equation $p(x)=-1$ than
roots $x_j$. The additional solutions are called {\it holes}. For an
illustration of a typical case see Fig~\ref{fig:gs}.

The corresponding distributions of roots and holes remain discrete
even in the limit of infinite Trotter number $N\to\infty$. The method
of studying these equations was developed in \cite{AK} and results
into two equations for two auxiliary functions
${\mathfrak{a}}(x):=1/p(x-i\gamma/2)$ and
$\overline{{\mathfrak{a}}}(x):=p(x+i\gamma/2)$ (corresponding to 
$S=1/2$ spinon and antispinon dressed energy functions)
\begin{equation}
\ln{\mathfrak{a}}(x) 
= -\frac{v\beta}{\cosh\frac{\pi}{\gamma} x}
+\frac{\pi\beta h}{2(\pi-\gamma)}
+[\kappa\ast_1\ln(1+{\mathfrak{a}})](x) 
- [\kappa\ast_2\ln(1+\overline{{\mathfrak{a}}})](x-
i\gamma +i\epsilon).\label{nliea}
\end{equation}
An analogous integral equation for $\overline{\mathfrak{a}}$ can be
obtained from the above one by use of the obvious identity
$\ln\overline{\mathfrak{a}}(x) = -\ln{\mathfrak{a}}(x +i\gamma)$
thus completing the non-linear integral equations.
%
%
\begin{figure}
\centering
\includegraphics[width=0.49\textwidth]{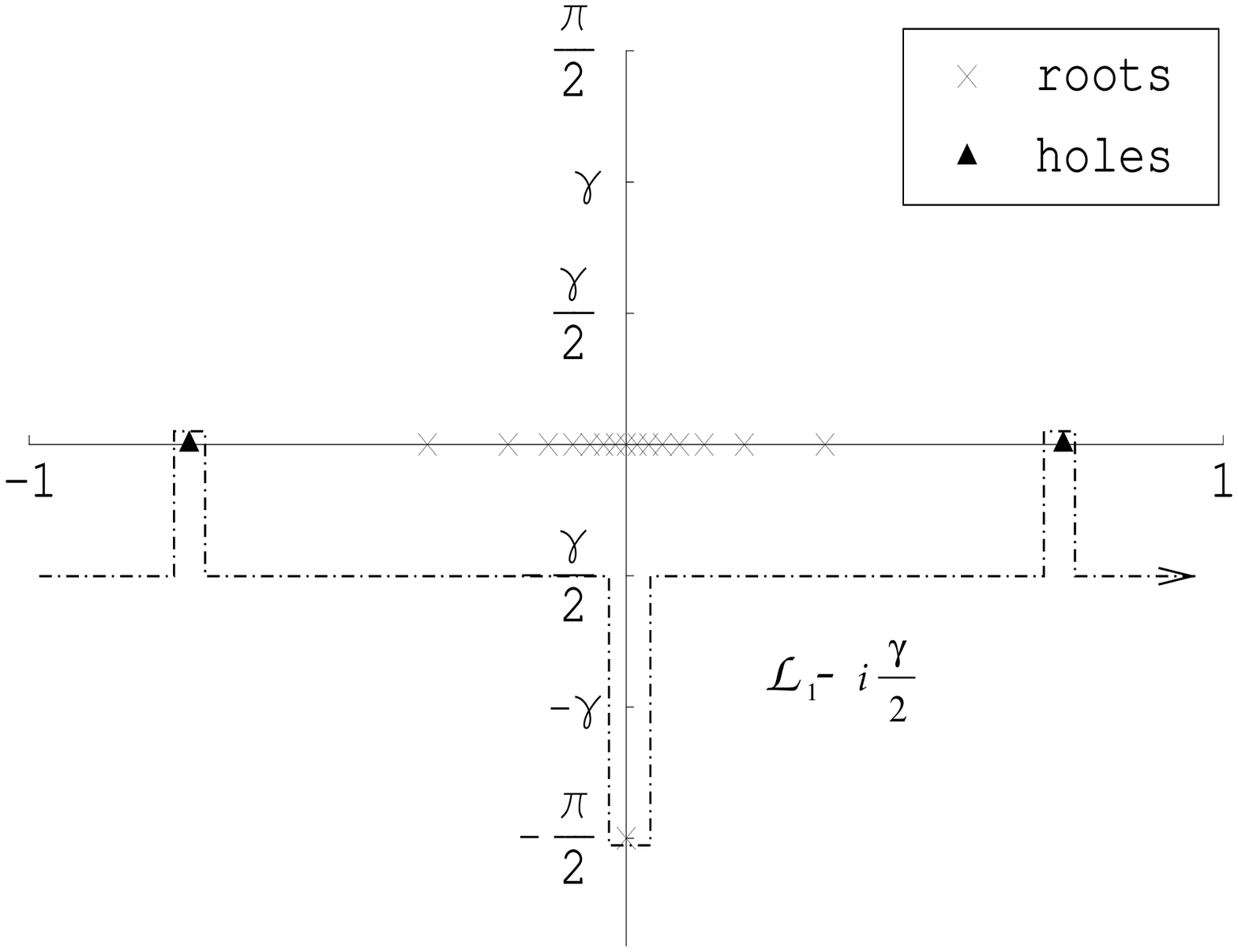}
\includegraphics[width=0.49\textwidth]{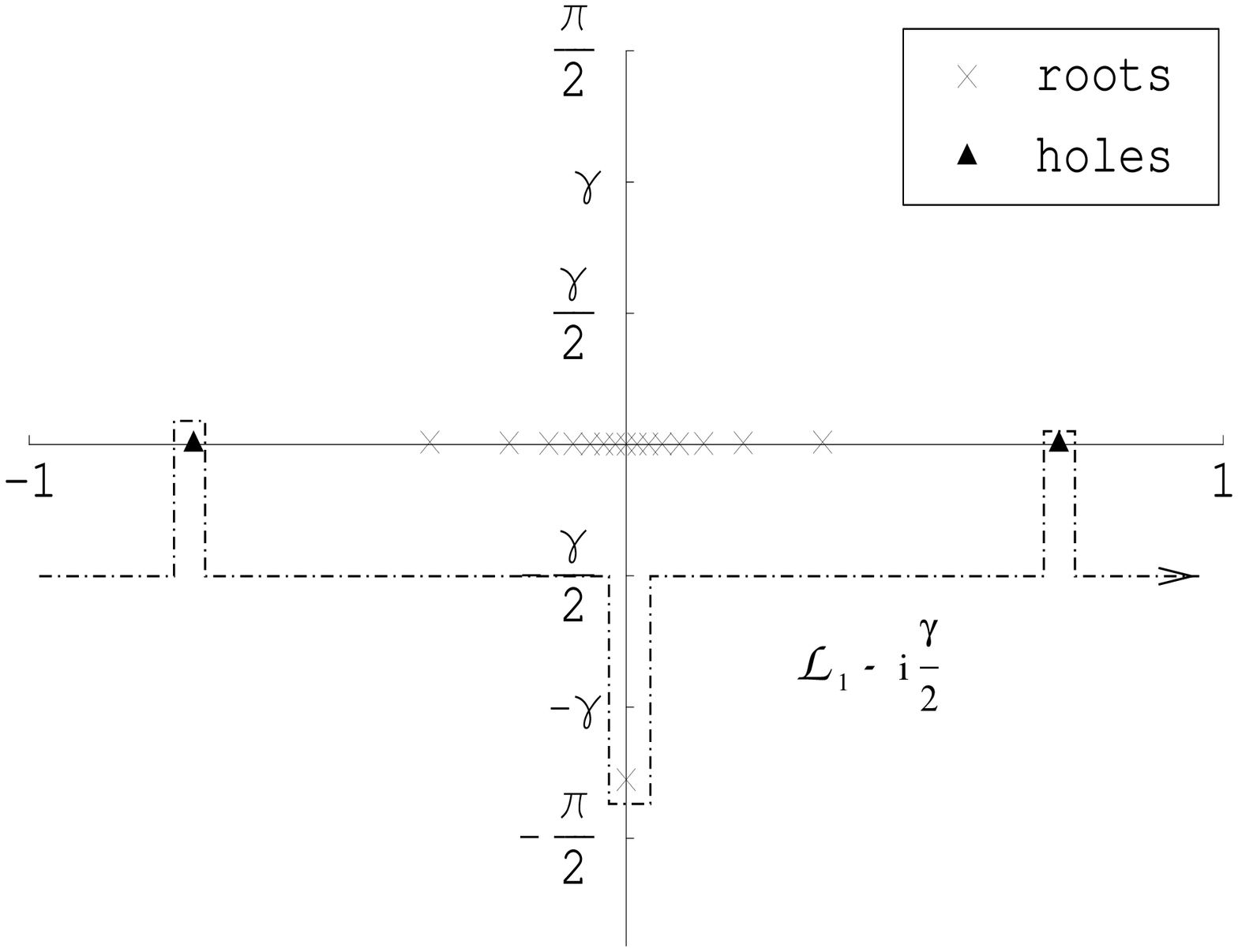}
\caption{Distribution of roots and holes for a 1-string solution for
anisotropy parameter $\gamma=\pi/3$, reciprocal temperature $\beta =
3.0$, and (a) $h = 0$ and (b) $h=0.6$.  Of all hole solutions only the
two new holes close to the real axis are shown. Note that for (a) the
1-string is situated exactly at $-i\pi/2$ and the holes lie on the
real axis. For positive magnetic field (b) the 1-string is shifted
upwards (see also Fig.~\ref{fig:crossover}).
The contour ${\mathcal{L}}_1-i\gamma/2$ is depicted by a
dashed-dotted line.}
\label{1fig300}
\end{figure}
\begin{figure}
\centering
\includegraphics[width=0.49\textwidth]{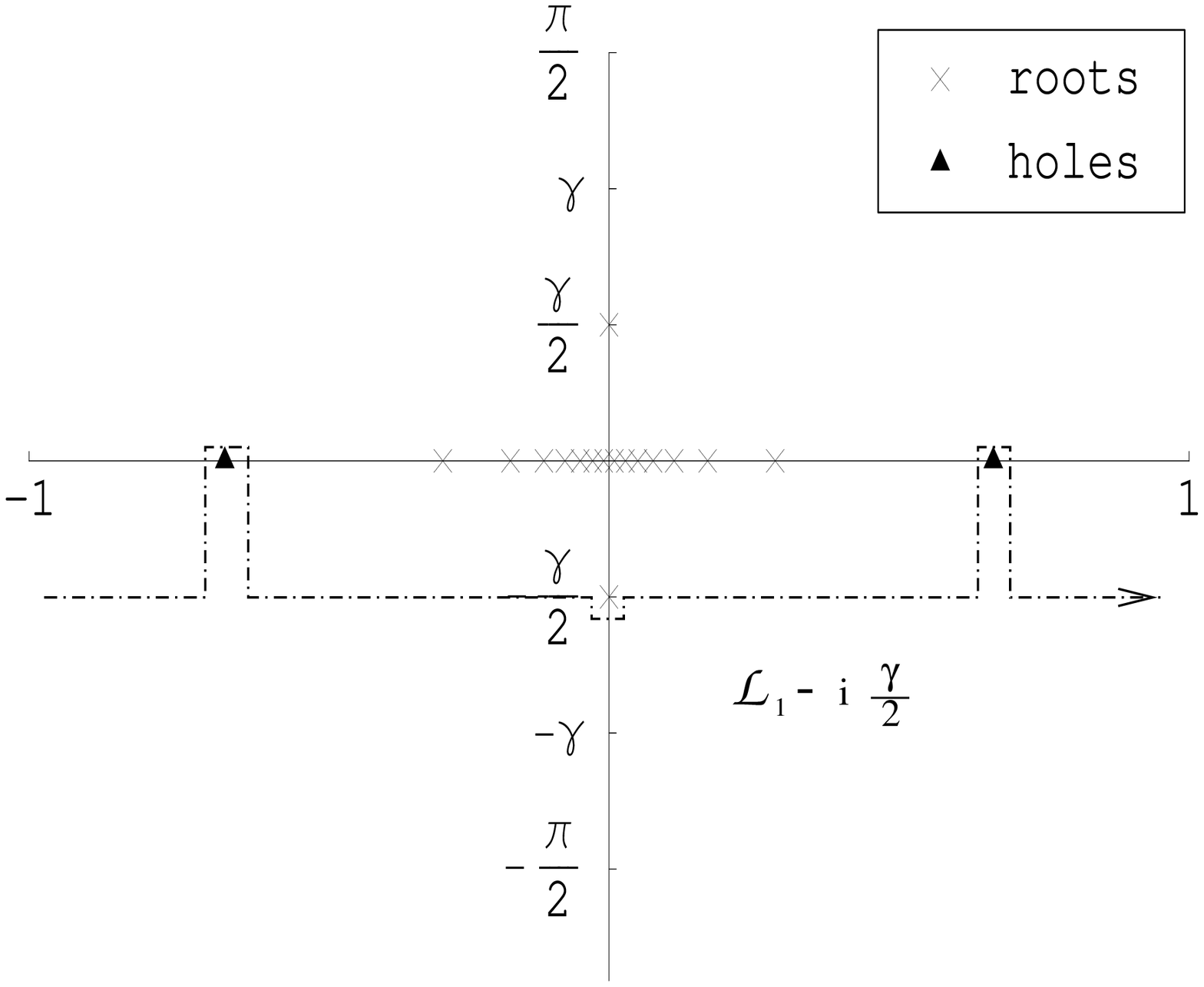}
\includegraphics[width=0.49\textwidth]{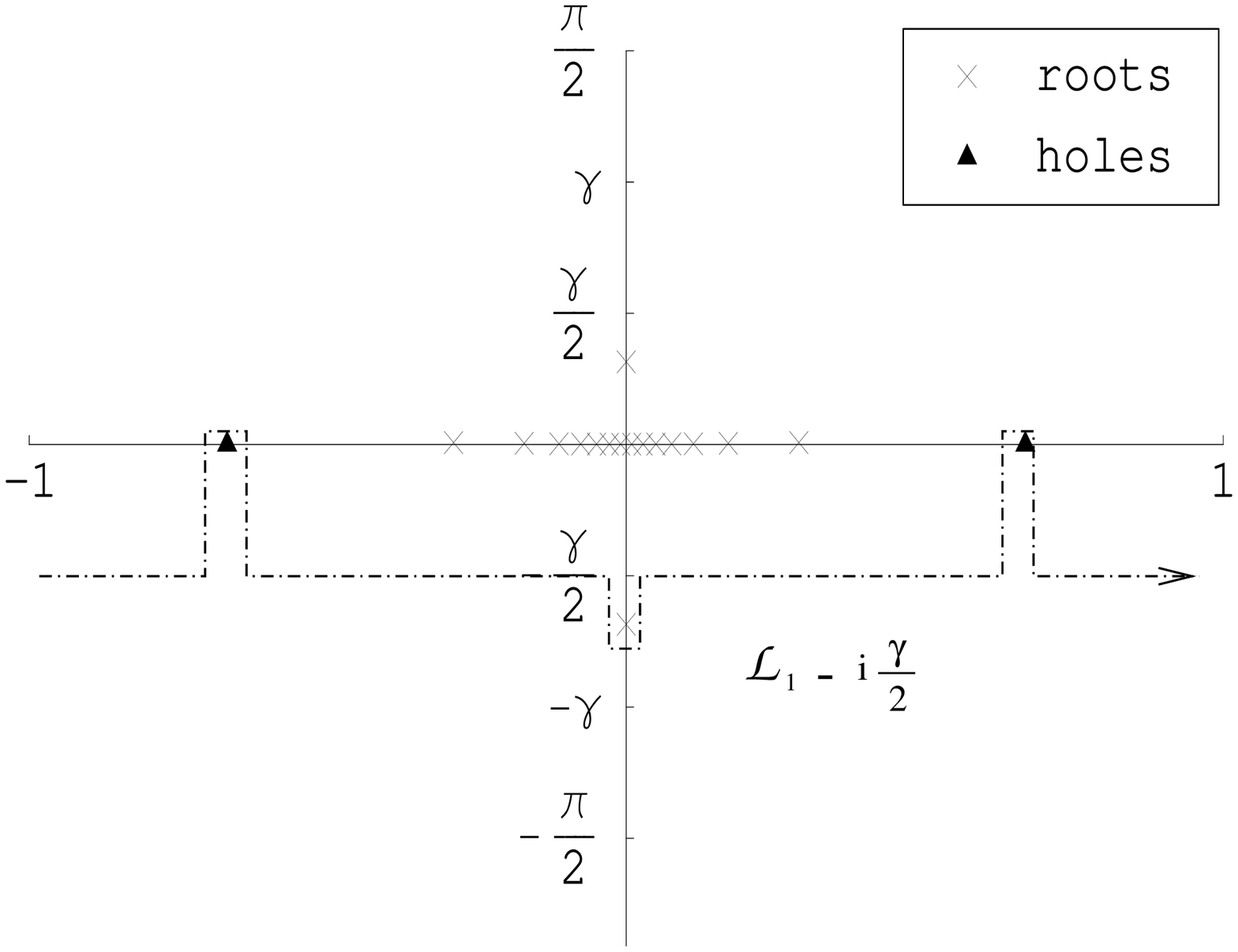}
\caption{Distribution of roots and holes for a 
2-string solution for anisotropy parameter $\gamma=\pi/3$, reciprocal
temperature $\beta = 3.0$, and (a) $h = 0$ and (b) $h=0.6$.  Only the
two new holes close to the real axis are shown. Note that for (a) the
2-string is symmetric with respect to the real axis and situated
approximately at $\pm i\gamma/2$; the holes lie on the real axis. For
positive magnetic field (b) the 2-string is shifted downwards (see also
Fig.~\ref{fig:crossover}). The contour
${\mathcal{L}}_1-i\gamma/2$ is depicted by a
dashed-dotted line.}
\label{2fig300}
\end{figure}
%
The integral kernel $\kappa(x)$ is defined by a Fourier integral
\begin{equation}
\kappa(x) :=
\frac{1}{2\pi}\int_{-\infty}^{\infty}
\frac{\sinh(\frac{\pi}{2}-\gamma)k\,\cos(kx)}{2\cosh\frac{\gamma}{2}k
\,\sinh\frac{\pi-\gamma}{2}k}\,\hbox{d}k.
\end{equation}
The symbol $\ast_{\mathcal{L}}$ denotes convolution
\begin{equation}
f\ast g(x) =
\int_{\mathcal{L}} f(x-y)g(y)dy 
\end{equation}
with a suitably defined integration contour ${\mathcal{L}}$. In
\refeq{nliea} the subscripts $\ast_1$ ($\ast_2$) refer to integration
paths ${\mathcal{L}}_1$ and ${\mathcal{L}}_2$ that extend from
$-\infty$ to $+\infty$ and lie below the distribution of numbers
$x_j+i\gamma/2$ and above $x_j-i\gamma/2$. An additional and last
requirement for these paths is that all roots $x_j$ be situated
between ${\mathcal{L}}_1-i\gamma/2$ and ${\mathcal{L}}_2+i\gamma/2$,
but no hole solution to the Bethe ansatz equation.

For the largest eigenvalue the distribution of roots and holes is
depicted in Fig.~\ref{fig:gs}. Here the paths ${\mathcal{L}}_1-i\gamma/2$ and 
${\mathcal{L}}_2+i\gamma/2$ are just straight lines with imaginary parts
$\mp\gamma/2$. Hence, ${\mathcal{L}}_{1,2}$ coincide with the real axis.

The cases of next-leading excitations which are of 1-string type (root
$y_0$ in the lower half plane) and 2-string type (upper and lower
roots $y_+$ and $y_-$
separated by approximately $i\gamma$) are shown in
Figs.~\ref{1fig300}, \ref{2fig300}. For the excited states the
contours of integration are deformed.  Only the path
${\mathcal{L}}_1-i\gamma/2$ is shown explicitly,
${\mathcal{L}}_2+i\gamma/2$ is simpler as it is mostly following a
straight line with imaginary part $+\gamma/2$ with loop in counter clockwise
manner around $y_0+i\gamma$ and $y_-+i\gamma$, respectively.

The contours may be straightened and by use of Cauchy's theorem we
pick up residues that lead to additive contributions to the driving
terms in the non-linear integral equations \cite{AK}.  This form of
the NLIE is particularly useful for numerical treatments\cite{KMSS}.

Finally, the subsidiary conditions ${\mathfrak{a}}(y_0 +i\gamma/2)= -1$
and ${\mathfrak{a}}(y_\pm +i\gamma/2)= -1$ with $y_0$, $y_+$ and $y_-$
denoting the root of the 1-string and the upper and lower constituents
of the 2-string, respectively, yield the information on the positions
of the string parameters.

The eigenvalues of the QTM are expressed in terms of
${\mathfrak{a}}(x)$, $\overline{\mathfrak{a}}(x)$ 
\begin{equation}
  \label{lambda}
  \L=-\beta e_0 + \frac{1}{2\gamma}
\int\frac{\ln[(1+{\mathfrak{a}}(x))(1+\overline{\mathfrak{a}}(x))]}{\cosh
  \frac{\pi}{\gamma} x}dx,
\end{equation}
where $e_0$ is the groundstate energy for zero magnetic field and 
the integration contours are ${\mathcal{L}}_{1,2}$. However, due to 
certain analyticity properties of the integrand these contours 
can be simplified to just one contour along the real axis, surrounding the
numbers $\theta+i\gamma/2$ in clockwise manner where $\theta$ is any 
of the hole solutions to the Bethe ansatz equation close to the real axis. 

The range of validity (convergence) of the integral equation
\refeq{nliea} for ${\mathfrak{a}}(x)$ is the strip
Im$(x)\in[0,\gamma]$.  Sometimes it is necessary to extend the above
equation to the strip Im$(x)\in[-(\pi-\gamma),0]$, e.g. for calculating 
${\mathfrak{a}}(y_0 +i\gamma/2)= -1$ with $y_0$ in the lower half plane, see
Figs.~\ref{1fig300}. 
One version of such an expression for the 1-string pattern is
\begin{eqnarray}
\ln{\mathfrak{a}}(x) 
&=& \frac{\pi\beta h}{\pi-\gamma}
-[r\ast\ln(1+{\mathfrak{a}})](x) 
+[r\ast\ln(1+\overline{{\mathfrak{a}}})](x-i\gamma)\cr
&&+\log\left[
\frac{\sh(x-y_0-i3\gamma/2)}{\sh(x-y_0+i\gamma/2)}
\prod_{j=1,2}
\frac{\sh(x-\theta_j+i\gamma/2)}{\sh(x-\theta_j-i\gamma/2)}
\right].\label{nle2}
\end{eqnarray}
where
\begin{equation}
r(x)=\frac{i}{2(\pi-\gamma)}\left(\cth x-\cth (x+i\gamma)\right).
\end{equation}
Note that the above expression for ${\mathfrak{a}}(x)$ is valid for $x$
below the real axis, however ${\mathfrak{a}}$, $\overline{{\mathfrak{a}}}$
on the right hand side are evaluated for strictly real arguments, i.e.
$\ast$ denotes standard convolution with the integration contour along
the real axis. In particular, there are no deformations of the contour as
discussed above. They have been removed by use of Cauchy's theorem. In turn,
the rapidities $\theta_1$, $\theta_2$ and the 1-string parameter $y_0$
show up explicitly. Note that the corresponding expression for the 2-string
pattern is obtained by simply exchanging $y_0$ with $y_-$ (the lower part
of the 2-string). In this case, the upper part $y_+$ does not appear
explicitly.
The details will be presented in \cite{KMSS}.
%

%
%
%
\begin{figure}[t]
  \begin{center} \leavevmode
    \includegraphics[width=0.85\textwidth]{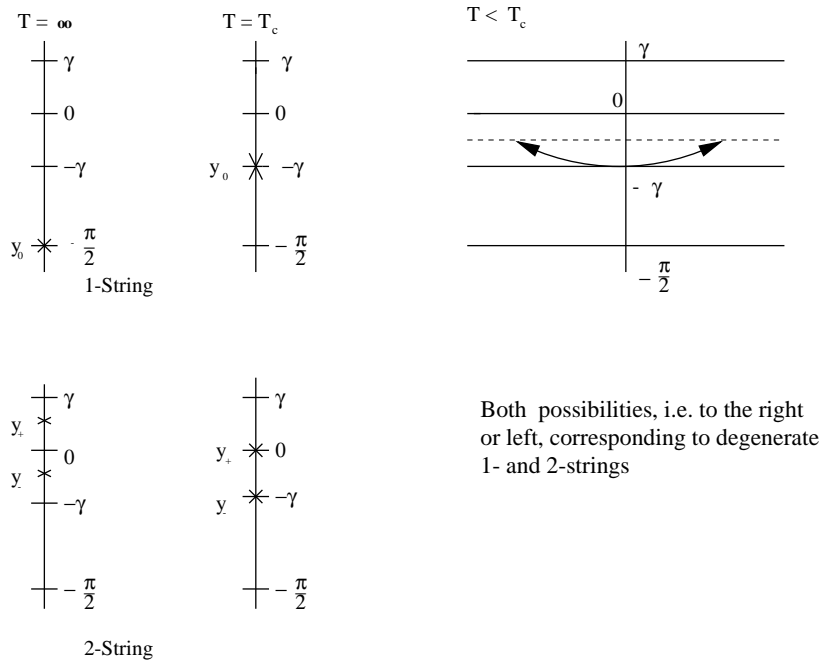}
    \caption{Schematic depiction of the crossover scenario.  At the
    left we show for the 1-string and 2-string solutions how the
    strings move away from their high temperature positions upon a
    decrease of $T$. At some well-defined temperature $T_c$ both solutions
    become degenerate. A further decrease of temperature forces the
    1-string and the lower part of the 2-string to move away from the
    imaginary axis.  As a result we have two complex conjugate
    solutions with complex eigenvalues.}
\label{fig:crossover} 
\end{center}
\end{figure}

\section{Crossover scenario and the particle-hole picture}
\label{sec:cross}

\begin{figure}[h]
 \includegraphics[width=0.49\textwidth]{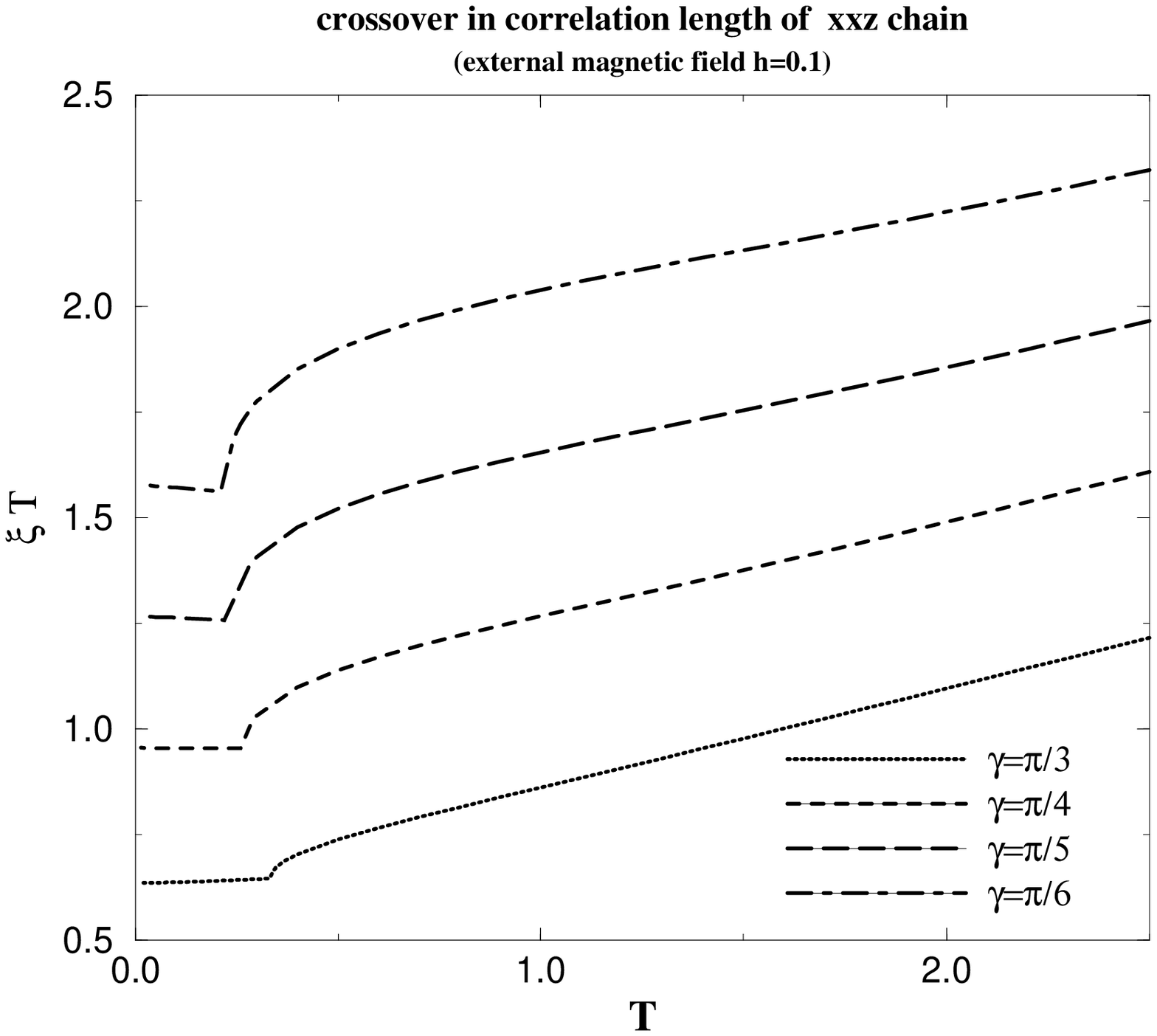}
 \includegraphics[width=0.49\textwidth]{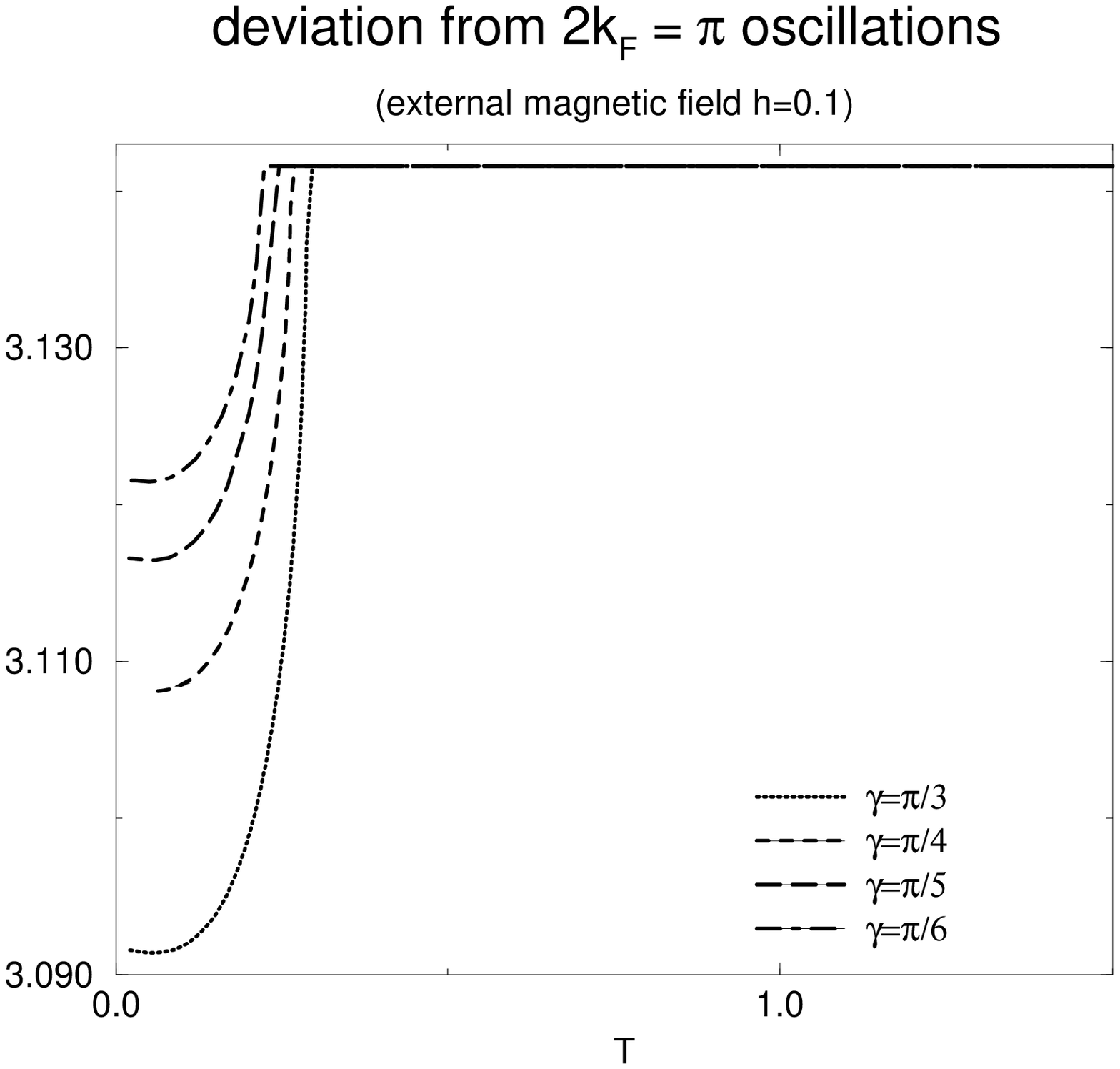}
\caption{The $XXZ$ chain for finite field $h=0.1$ and
several values of the anisotropy parameter $\gamma$. Temperature
dependence of: (a) longitudinal correlation length $\xi$ times $T$,
(b) Fermi-momentum. The crossover temperature is monotonously 
decreasing for decreasing $\gamma$ with a seemingly
finite limiting value for $\gamma\to 0$.}
\label{fig:corr}
\end{figure}

From the next-largest eigenvalue(s) $\Lambda_1$ of the QTM the
asymptotic behaviour of the correlation functions is determined.  
If $\Lambda_1$ is real and positive (negative), the decay is purely 
exponential without (with) sublattice oscillations. In general 
$\Lambda_1$ may be complex and the asymptotics are characterized by 
a finite correlation length $\xi$ and 
Fermi-momentum $k_F$ defined by
$$
\frac{\Lambda_1}{\Lambda_0} = \e^{-1/\xi \pm i2 k_F},
$$
We have numerically solved the nonlinear integral equations 
for the next-leading eigenvalue(s) and plot the results
in Figs.~\ref{fig:crossover}, \ref{fig:corr}. 

Summarizing our results we see that the previously drawn picture of
excitations of distinctly 1-string and 2-string type is valid only for
$h=0$ or, if $h>0$ for sufficiently high $T$. For this case the
1-string solution dominates and the eigenvalue is real because of the
left-right symmetry of the Bethe ansatz pattern.  The 2-string
solution is subdominant. Both strings lie on the imaginary axis where
for $T=\infty$ the 1-string is located at $-i\pi/2$ and the 2-string is
symmetric with respect to the real axis.  For decreasing temperature
we observe a characteristic motion of the 1-string upwards and the
2-string downwards along the imaginary axis,
cf. Fig.~\ref{fig:crossover}.  This motion continues until the root of
the 1-string ($y_0$) and the lower root of the 2-string ($y_-$) take
identical values. The corresponding temperature defines $T_c$. For
lower temperatures a horizontal motion sets in. The considered root
previously on the imaginary axis develops a non-vanishing real part
which may be positive or negative, see Fig.~\ref{fig:crossover}. The
corresponding Bethe ansatz patterns are related by reflection at the
imaginary axis, the corresponding eigenvalues are complex
conjugate. There is no longer any qualitative distinction like for
temperatures higher than $T_c$.

The reason for this crossover may be understood qualitatively in the
following way. The subsidiary condition for the 1-string $y_0$ (or for
the 2-string with $y_0$ replaced by $y_-$) is
${\mathfrak{a}}(y_0+i\gamma/2)=-1$ which yields due to \refeq{nle2}
\begin{eqnarray}
&&\frac{\pi\beta h}{\pi-\gamma}
+\log\left[
\frac{\sh(y_0-\theta_1+i\gamma)}{\sh(y_0-\theta_1)}
\frac{\sh(y_0-\theta_2+i\gamma)}{\sh(y_0-\theta_2)}
\right]\cr
&&=\hbox{``integral expressions''}.\label{nle2app}
\end{eqnarray}
Unfortunately, the right hand side has to be evaluated numerically. It
turns out to be of order $O(\beta h)$, but smaller than the first term
on the left hand side. In any case, the equation \refeq{nle2app} for
$y_0$ has two inequivalent solutions. These solutions are purely
imaginary and different
for small $\beta h$ ($\leftrightarrow$ distinct 1- and 2-strings);
they have same imaginary part but
non-vanishing real parts with opposite signs for large $\beta h$
($\leftrightarrow$ degenerate 1- and 2-strings). The crossover is typically
associated with square root singularities. The precise value of $T_c$
can only be calculated numerically.

The results for the correlation length and Fermi momentum are shown in
Fig.~\ref{fig:corr}. Most strikingly we see a non-analytic temperature
dependence for non-vanishing external magnetic fields at the
well-defined crossover temperature $T_c$!  The singularity at $T_c$ is
of square root type. Most significant is the non-analytic behaviour of
the Fermi momentum. At low temperature the oscillations are
incommensurate and at zero temperature the Fermi momentum $k_F$ and
magnetization $m$ are strictly related by $k_F=(1/2-m)\pi$, a relation
which ceases to hold at elevated temperatures.  At sufficiently high
temperatures the oscillations are commensurate with $k_F=\pi/2$.  The
loss of the left-right symmetry in the Bethe ansatz patterns at low
temperatures is the reason for the incommensurability of $k_F$ and the
non-analytic behavior of the correlation length.

\begin{figure}[htbp]
  \begin{center} \leavevmode
    \includegraphics[width=0.8\textwidth]{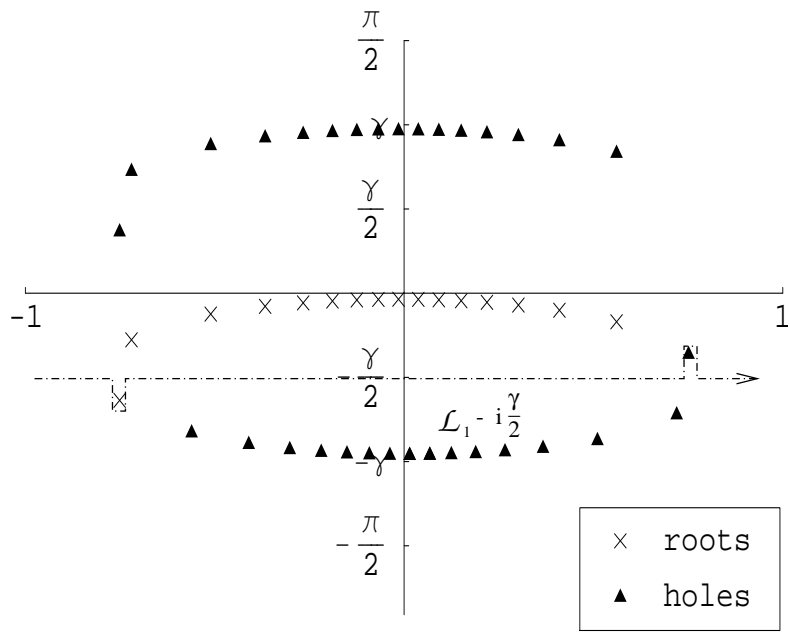}
    \caption{Bethe ansatz pattern at low temperature ($h=0.6$ and $\beta=12.0$)
    corresponding to a 1-string (or 2-string) solution at sufficiently
    high temperature.}  
\label{fig:p-h} 
\end{center}
\end{figure}

We note that temperature and magnetic field act in roughly opposite
ways which can be inferred in part from their appearance in the
combination $h/T$ in the NLIE. At sufficiently low temperatures the
distinguishing characters of the 1-string and 2-string disappear
completely and a root-hole picture emerges as shown in
Fig~\ref{fig:p-h}. Here the relevant modifications of the excited
state of the QTM in comparison to the ground state (see
Fig.~\ref{fig:gs}) can be characterized as a rearrangement of the
roots and holes on an ellipsoidal curve. Clearly, the same
constructions and terminology as used for the excitations of Fermi
systems applies.  In this finite field case ($h>0$), the analytic
treatment of the low-temperature asymptotics not only confirms the
predictions of CFT, but also recovers completely the dressed charge
formalism as will be shown in\cite{KMSS}.
\section*{Acknowledgments}
The authors like to acknowledge valuable discussions with 
F. E\ss ler, H. Frahm, F. G\"ohmann, M. Inoue, K. Sakai, and J. Suzuki.
A.K. acknowledges financial support by the {\it Deutsche
Forschungsgemeinschaft} under grant No.  Kl~645/3-2 and support by the
research program of the Sonderforschungsbereich 341,
K\"oln-Aachen-J\"ulich.

\end{document}